\documentclass[
    aps,
    prd,
    amssymb,
    superscriptaddress,
    floatfix,
    nofootinbib,
    reprint
]{revtex4-2}

\usepackage{amsmath}
\usepackage{amsfonts}
\usepackage{bm}
\usepackage{braket}
\usepackage{graphicx}
\usepackage{booktabs}
\usepackage{multirow}
\usepackage{dcolumn}
\usepackage{enumitem}
\usepackage{lineno}
\usepackage[dvipsnames,x11names]{xcolor}
\usepackage[normalem]{ulem}
\usepackage{soul}
\usepackage{hyperref}
\hypersetup{
    colorlinks=true,
    citecolor=Green,
    urlcolor=Blue,
    linkcolor=Blue
}
\usepackage{comment}
\usepackage{mfirstuc}
\usepackage{orcidlink}

\usepackage[T1]{fontenc}
\usepackage{microtype}
\usepackage[vvarbb,smallerops,lining]{newtx}

\usepackage[compact]{titlesec}

\titleformat{\section}
    {\centering\small\bfseries\MakeUppercase}
    {\noindent\thesection.}
    {2pt}{}

\titleformat{\subsection}
    {\centering\small\bfseries}
    {\noindent\thesubsection.}
    {2pt}{}

\titlespacing\section{0pt}{18pt plus 1pt minus 1pt}{8pt plus 1pt minus 1pt}
\titlespacing\subsection{0pt}{12pt plus 1pt minus 1pt}{8pt plus 1pt minus 1pt}

\newcommand{\tstrut}{\rule{0pt}{2.6ex}}

\makeatletter

\newdimen\tov@rt
\newcommand{\tovgap}{2}

\newcommand{\tov@setrule}[1]{%
    \ifx#1\scriptscriptstyle
        \tov@rt=\fontdimen8\scriptscriptfont3
    \else\ifx#1\scriptstyle
        \tov@rt=\fontdimen8\scriptfont3
    \else
        \tov@rt=\fontdimen8\textfont3
    \fi\fi
}

\newcommand{\tov@build}[2]{%
    \tov@setrule{#1}
    \sbox0{$\m@th#1#2$}%
    \vbox{%
        \offinterlineskip
        \kern\tov@rt
        \hrule height\tov@rt
        \kern\tovgap\tov@rt
        \box0
    }
}

\newcommand{\tightoverline}[1]{\mathpalette\tov@build{#1}}

\makeatother

\newcommand{\olsi}[1]{\,\tightoverline{\!#1}}

\begin{document}

 \title{\boldmath The $D\pi$ and $D^*\pi$ femtoscopy puzzle }
 
\newcommand{\ific}{\affiliation{\small%
Instituto de F\'isica Corpuscular, Centro Mixto Universidad de Valencia-CSIC, \\
Institutos de Investigaci\'on de Paterna, Apartado 22085, E-46071, Valencia, Spain}}

\newcommand{\teoUV}{\affiliation{\small%
Departamento de F\'\i sica Te\'orica and IFIC, Centro Mixto Universidad de Valencia-CSIC,
Institutos de Investigaci\'on de Paterna, Aptdo. 22085, E-46071 Valencia, Spain}}

\author{Pablo Encarnación\orcidlink{0009-0005-0749-3885}}
\email{Pablo.Encarnacion@ific.uv.es}
\teoUV

\author{Miguel Albaladejo\orcidlink{0000-0001-7340-9235}}
\email{Miguel.Albaladejo@ific.uv.es}
\ific 

\author{Albert Feijoo\orcidlink{0000-0002-8580-802X}}
\email{Eduardo.Feijoo@ific.uv.es}
\teoUV

\author{Juan Nieves\orcidlink{0000-0002-2518-4606}}
\email{Juan.M.Nieves@ific.uv.es}
\ific

\renewcommand{\abstractname}{\vspace{20pt}Abstract}

\begin{abstract}
 We present a relativistic momentum-space framework for femtoscopic correlation functions including strong and Coulomb interactions in coupled channels. The approach extends the Vincent--Phatak prescription to the three-dimensional Bethe--Salpeter equation, combining a fully relativistic treatment with the correct Coulomb asymptotics through wave-function matching. We apply it to the $D\pi$ system using a coupled-channel heavy-meson chiral interaction that reproduces the established two-pole structure of the $D_0^*(2300)$.  Our correlation functions agree with previous theoretical calculations but differ significantly from those extracted by the ALICE Collaboration. We trace this tension to the assumption made in the experimental analysis that the genuine correlation function reaches unity above a relative momentum of $100$ MeV, whereas this region remains affected by the lower $D_0^*(2300)$ pole. Repeating the extraction over different fitting windows reveals a strong dependence of the extracted signal on the chosen range, demonstrating that the resulting $D\pi$ scattering parameters are not robust and motivating a dedicated reanalysis of the experimental data. We extend the analysis to the $D^*\pi$ system, where the same issue arises in close connection with heavy-quark spin symmetry.  Together with the unresolved tension between theoretical calculations and the ALICE $K\pi$ femtoscopic measurements, this highlights the need to clarify the origin of this puzzle, with the aim of establishing femtoscopy as a reliable quantitative tool for precision studies of low-energy hadron interactions. 
\end{abstract}

\maketitle

\setcounter{tocdepth}{3}

\section{Introduction}\label{sec:introduction}

The advent of hadron femtoscopy has provided a complementary technique for accessing hadron--hadron interactions through momentum correlations measured at small relative momenta in high-multiplicity collisions. This approach offers unique information on systems that are challenging to investigate through conventional scattering experiments and, in some sectors, represents the only available source of experimental information. The systems studied with hadron femtoscopy span the strange and charm sectors, as well as systems involving vector mesons \cite{ALICE:2019hdt,ALICE:2023wjz,ALICE:2022enj,ALICE:2021cpv,ALICE:2025flv}. Correlation functions (CFs) at low relative momenta are particularly sensitive to near-threshold dynamics, enabling the extraction of scattering parameters. Since the measured CFs are governed by final-state interactions, their theoretical description requires a consistent treatment of both short-range strong and long-range Coulomb forces.

While Coulomb effects are well established in coordinate-space approaches, their implementation in momentum-space coupled-channel calculations is considerably more challenging, particularly in relativistic formulations based on three-dimensional reductions of the Bethe--Salpeter equation (BSE), where the strong interaction is generally non-local. A momentum-space treatment based on the Vincent--Phatak prescription \cite{Vincent:1974zz} was recently developed for the single-channel case \cite{Encarnacion:2026iur}. Here, we extend this framework to relativistic coupled-channel dynamics, providing the full wave functions required to calculate femtoscopic CFs within the Koonin--Pratt formalism. The method is general and can be combined with arbitrary coupled-channel interaction models.

As an application, we investigate the $D\pi$ system, whose interaction has been the subject of a long-standing controversy since the ALICE Collaboration made its femtoscopic measurement publicly available~\cite{ALICE:2024bhk}. In particular, the analysis found a residual strong interaction to be consistent with zero, a result that appears to be in tension with theoretical descriptions based on coupled-channel $(D\pi,D_s\bar K,D\eta)$ heavy-meson chiral dynamics at next-to-leading order (NLO), constrained by LQCD calculations of selected scattering lengths~\cite{Guo:2008gp,Liu:2012zya}. These approaches predict a sizable $D\pi$ interaction and a characteristic two-pole structure for the $D_0^*(2300)$, first signaled in Ref.~\cite{Guo:2006fu} and subsequently established in Refs.~\cite{Albaladejo:2016lbb,Du:2017zvv} (see also Refs.~\cite{Guo:2018tjx, Meissner:2020khl, Messchendorp:2025men, Dai:2026fkg} and the review on {\it Heavy non $q\bar q$-Mesons} in the PDG~\cite{ParticleDataGroup:2026mpi}).  Notably, Ref.~\cite{Albaladejo:2016lbb} showed that the NLO chiral interaction reproduces the finite-volume energy levels of Ref.~\cite{Moir:2016srx} without fitting any additional parameters. In the SU(3) limit, the lower pole belongs to an $\overline{\mathbf{3}}$ multiplet, completed by the $D_{s0}^*(2317)$, whereas the higher pole belongs to the $\mathbf{6}$ and contains hidden-strangeness components. Further support for this picture was provided in Refs.~\cite{Du:2017zvv,Du:2019oki}, where the chiral amplitudes constrained in Ref.~\cite{Albaladejo:2016lbb} were found to be fully consistent with high-quality LHCb data on the $B^- \to D^+\pi^-\pi^-$ and $B_s^0 \to \bar{D}^0K^-\pi^+$ decay channels~\cite{LHCb:2016lxy,LHCb:2014ioa}. Subsequent studies have further explored the implications of chiral symmetry for scalar charmed mesons~\cite{Du:2019oki,Du:2020pui}. Ref.~\cite{Asokan:2022usm}, for instance, examined the compatibility of the two-pole structure with the LQCD study of $D\pi$ scattering at different pion masses~\cite{Moir:2016srx}, where only one pole was identified in the $D_0^*$ channel. A subsequent LQCD study without coupled channels~\cite{Gayer:2021xzv} found properties consistent with those of the lower pole. Moreover, LQCD information on the $D\to\pi\ell\nu$ transition was used in Ref.~\cite{Flynn:2007ki} to predict an $I=1/2$ $S$-wave resonance at $(2.2\pm0.1)$~GeV, compatible with the lower pole found in unitarized chiral approaches, while Ref.~\cite{Du:2025beb} investigated the role of the $D_0^*(2100)$ in $B\to D\pi\ell\nu$ decays. More recently, Ref.~\cite{Zhuang:2026lta} reported a two-pole structure for the $D_0^*(2300)$, with particular emphasis on its interpretation in the SU(3) limit. Taken together, these results point to a consistent theoretical picture, supported by both LQCD calculations and independent analyses of $B^-$ and $B_s^0$ decays, and therefore call for a careful examination of the origin of the apparent tension with the ALICE femtoscopic measurement. In this work, we revisit the theoretical calculation and confront its predictions with the ALICE result, while critically reassessing the experimental analysis to identify the source of the discrepancy and determine whether it can provide a plausible explanation for the observed absence of a sizable $D\pi$ strong interaction.

Using our relativistic formalism with a coupled-channel heavy-meson chiral interaction, we first show that the resulting $D\pi$ CFs are consistent with previous theoretical calculations, ruling out the treatment of Coulomb effects as the origin of the discrepancy with experiment. We then identify a possible source of the apparent tension in the assumption underlying the experimental extraction of the genuine CF\, namely, that it reaches unity above a relative momentum of $100$ MeV. In fact, this region still contains significant strong-interaction effects associated with the lower $D_0^*(2300)$ pole, which can bias the extracted CF. Repeating the extraction for different fitting windows reveals a pronounced dependence on the chosen range, suggesting that the ALICE conclusion of a negligible $D\pi$ interaction is not robust against this assumption.

We extend our analysis to the $D^*\pi$ system, for which the discussion closely parallels that of the $D\pi$ case owing to heavy-quark spin symmetry (HQSS). The LHCb analyses of the $B^- \to D^{*+}\pi^-\pi^-$ and $B^+ \to D_s^+D^{*-}\pi^+$ decays~\cite{LHCb:2019juy,LHCb:2024vhs} provide evidence for sizable low-energy $D^*\pi$ final-state interactions in the $D^{*+}\pi^-$ and $D^{*-}\pi^+$ channels, respectively. These findings are consistent with the unitarized chiral framework of Refs.~\cite{Guo:2008gp,Liu:2012zya, Albaladejo:2016lbb,Du:2017zvv}, but at odds with the $D^*\pi$ scattering lengths extracted from the femtoscopy analysis of Ref.~\cite{ALICE:2024bhk}.

\section{Femtoscopy formalism}\label{sec:formalism}

The CF for a pair of hadrons is defined through the 
Koonin--Pratt (KP) formula \cite{Bauer:1992ffu, Lisa:2005dd, Ohnishi:2016elb,Fabbietti:2020bfg,Vidana:2023olz, Albaladejo:2024lam},
\begin{equation}    \label{eq:KPformula}
    C(\vec k) = \int d^3 \vec r \,S(\vec r)\, |\Psi(\vec k,\vec r)|^2,
\end{equation}
where $\vec k$ denotes the relative momentum of the pair in the center-of-mass (CM) frame. The source function $S(\vec r)$ describes the probability distribution for emitting the two particles with relative separation $\vec r$. Moreover, $\Psi(\vec k,\vec r)$ is the relative wave function that incorporates both strong and Coulomb interactions. Since the strong interaction considered here is restricted to the $\ell=0$ partial wave, only the S-wave component of the wave function is modified, such that\footnote{We follow here the conventions and normalizations of Ref.~\cite{Albaladejo:2025kuv}, where $\langle \vec p\,' | \vec p\,\rangle= (2\pi)^3\delta^3(\vec p - \vec p\,')$, similarly $\langle \vec r\,' | \vec r\,\rangle= \delta^3(\vec r - \vec r\,')$, and hence $\langle\vec r\, | \vec p\,\rangle =e^{i\vec p\cdot\vec r}$.}
\begin{subequations}    \label{eq:coulombanalyticwf}
\begin{align}
\Psi(\vec k,\vec r) &  = \Psi^C(\vec k,\vec r) - \psi^C_{0}(k,r) + \psi_0(k,r)\,,\\
\!\Psi^C(\vec k,\vec r) &  =  e^{-\frac{\pi\eta}{2}} \Gamma(1\!-\!i\eta)\, M(i\eta,1;-ikr\!-\!i\vec k \cdot \vec r)\,e^{i\vec k \cdot \vec r} \,,
\end{align}
\end{subequations}
where $\Psi^C(\vec k,\vec r)$ denotes the full Coulomb wave function, $\psi^C_0(k,r)$ its S-wave projection, and $\psi_0(k,r)$ the $\ell=0$ partial-wave function including both strong and Coulomb interactions. We use the relativistic Coulomb parameter $\eta=Z_1Z_2\alpha E_s/k$, where $E_s=(s-m_1^2-m_2^2)/(2\sqrt{s})$ (see Sect.~\ref{sec:rcp} of the Appendix), with $\sqrt{s}$ the total CM energy of the system, $Z_{1,2}$ the corresponding electric charges in units of the proton charge, and $\alpha\simeq1/137$ the fine-structure constant. The function $M(a,b;z)={}_1F_1(a,b;z)$ denotes the confluent hypergeometric function.

For coupled channels, the CF of channel $i$ becomes \cite{Haidenbauer:2018jvl}
\begin{eqnarray}    \label{eq:swaveKP}
    C_i(k) = C^C_i(k) &-& \! \int \!\!d^3r\,S(r) |\psi^C_{0,i}(k,r)|^2 \nonumber \\
    &+& \! \int \!\!d^3r\,S(r) \sum_{j} w_j |\psi_0^{i\leftarrow j}(k,r)|^2 ,
\end{eqnarray}
where $C^C_i(k)=\int d^3\vec r\,S(r)\,|\Psi^C_i(\vec k,\vec r)|^2$ is the CF generated solely by the Coulomb interaction and $w_j$ are the channel production weights, normalized to $w_i=1$ for the observed channel.

The coupled-channel $S$-wave functions are obtained from the half-off-shell $T$-matrix through
\begin{eqnarray} \label{eq:bethesalpeterequationwavefunction}
    && \!\! [\psi_0^{i\leftarrow j}(k,r)]^* = j_0(kr)\delta_{ij} +  \\
    && \!\! \mathcal K_{ij} \int \!\! \frac{d^3 q}{(2\pi)^3}
    \frac{\omega_{1j}(q) \!+\! \omega_{2j}(q)}
    {2\omega_{1j}(q)\omega_{2j}(q)}
    \frac{T_{ij}(k,q;s)j_0(qr)}
    {s \!-\! (\omega_{1j}(q)+\omega_{2j}(q))^2  \!+\! i\epsilon} \,, \nonumber
\end{eqnarray}
with the boundary condition of channel $i$ being the observed (outgoing) channel. Here, $\mathcal K_{ij} = \sqrt{\omega_{1j}\omega_{2j}/\omega_{1i}\omega_{2i}}$, ensures the correct non-relativistic limit of the wave functions (see Sect.~\ref{sec:kij} of the Appendix) and  $\omega_{aj}(q) = \sqrt{q^2 + m_{aj}^2}$ is the energy of the $a$-th particle in channel $j$. 

The amplitudes $T_{ji}$ (transition from the state $i$ with momentum $p$, to the state $j$ with momentum $p'$) are obtained by solving the coupled-channel BSE
\begin{eqnarray} \label{eq:bethesalpeterequation}
    &&T_{ji}(p',p; s) = V_{ji}(p',p) +  \\
    && \sum_l \! \int \!\! \frac{d^3 q}{(2\pi)^3}
    \frac{\omega_{1l}(q) \!+\! \omega_{2l}(q)}
    {2\omega_{1l}(q)\omega_{2l}(q)}
    \frac{T_{jl}(p',q;s)V_{li}(q,p)}
    {s \!-\! (\omega_{1l}(q)+\omega_{2l}(q))^2 \!+\! i\epsilon} \,.
    \nonumber
\end{eqnarray}

\section{Strong interaction}

We study the $D\pi$ system in the physical basis, in the charge sector $Q=0$ comprising the coupled channels $D^+\pi^-$, $D^0\pi^0$, $D_s^+K^-$, and $D^0\eta$, as well as the $D^+\pi^+$ channel with $Q=2$.
\begin{table*}[t!]
    \centering
    \begin{tabular}{ccccccc}
        \hline
        $f_\pi$ [MeV] & $h_0$ & $h_1$ & $h_{24}$ & $h_4$ & $h_{35}$ & $h_5$ \\
        \hline
        \tstrut
        $92.21$ & $0.014$ & $0.42$ & $-0.10_{-0.06}^{+0.05}$ &
        $-0.32_{-0.34}^{+0.35}/\bar M_D^2$ &
        $0.25_{-0.13}^{+0.13}$ &
        $-1.88_{-0.61}^{+0.63}/\bar M_D^2$
        \tstrut\\[1mm]
        \hline
    \end{tabular}
    \caption{LECs of the potential in Eqs.~(\ref{eq:NLOstrongpotential}-\ref{eq:NLOH24H35}), taken from Table V and text of Ref.~\cite{Liu:2012zya}.}
    \label{tab:LECSfengkun}
\end{table*}

The strong interaction is described by the coupled-channel next-to-leading-order (NLO) heavy-meson chiral perturbation theory (HMChPT) potential of Refs.~\cite{Guo:2008gp,Liu:2012zya}, successfully applied in previous studies~\cite{Albaladejo:2016lbb,Du:2017zvv,Guo:2017jvc,Albaladejo:2018mhb,Du:2019oki,Du:2020pui},
\begin{eqnarray}    \label{eq:NLOstrongpotential}
    V_S(s,t,u) = \! && \!\!\!\! \frac{1}{f_\pi^2}\Big[\frac{C_{LO}}{4}(s-u) - 4C_0 h_0 + 2C_1 h_1 \nonumber \\
    &&- 2 C_{24}H_{24}(s,t,u) + 2C_{35}H_{35}(s,t,u) \Big],
\end{eqnarray}
where
$p_1$ $(p_2)$ and $p_3$ $(p_4)$ are the momenta of the heavy mesons
(Goldstone bosons) in the initial and final state, respectively.  $s=(p_1+p_2)^ 2$, $t=(p_1-p_3)^2$, $u=(p_1-p_4)^2$ are
the Mandelstam variables and  
\begin{subequations} \label{eq:NLOH24H35}
\begin{align}
    H_{24}(s,t,u) =& 2h_{24}(p_2\cdot p_4) + h_4[(p_1\cdot p_2)(p_3\cdot p_4) \nonumber \\
    &+ (p_1\cdot p_4)( p_2\cdot p_3) - 2\bar M_D^2 (p_2\cdot p_4)]\,, \\
    H_{35}(s,t,u) =& h_{35}(p_2\cdot p_4) + h_5[(p_1\cdot p_2)(p_3\cdot p_4) \nonumber \\
    &+ (p_1\cdot p_4)( p_2\cdot p_3) - 2\bar M_D^2 (p_2\cdot p_4)]\,, 
\end{align}
\end{subequations}
and $\bar M_D=(M_D^{phy}+M_{D_s}^{phy})/2$ is the average of the physical masses of the $D$ and $D_s$ mesons. The coefficient matrices $C_{LO}$, $C_0$, $C_1$, $C_{24}$ and $C_{35}$ are taken from Table VII of Appendix A of Ref.~\cite{Montana:2020vjg}. For the low-energy constants (LECs), we adopt the fit reported in Table V of Ref.~\cite{Liu:2012zya}. Their values are summarized in Table \ref{tab:LECSfengkun}. 
The analysis is restricted to the $\ell=0$ partial wave,
\begin{equation}    \label{eq:NLOstrongpotentialswave}
    V_S^{\ell=0}(p',p;p_{1,2,3,4}^0)=\frac{1}{2}\int_{-1}^1 d(\cos\theta_{pp'}) V_S(s,t,u) \,,
\end{equation}
where $p$ and $p'$ are the magnitudes of the incoming and outgoing center-of-mass (CM) three-momenta, respectively, and $\theta_{pp'}$ is the angle between them. 

We adopt the Blankenbecler--Sugar approach~\cite{Blankenbecler:1965gx}, also employed in Ref.~\cite{Torres-Rincon:2023qll}. The kernel matrix $V(p',p)$ in Eq.~\eqref{eq:bethesalpeterequation} is expressed in terms of three-momenta, whereas the strong interaction kernel in Eq.~\eqref{eq:NLOstrongpotential} depends on four-momenta. A prescription is therefore required to fix the zeroth components of the four-momenta in the three-dimensional reduction. For a given CM energy $\sqrt{s}$, we set the zeroth components of the external four-momenta to their corresponding on-shell values. For the intermediate off-shell momenta, their zeroth components are kept fixed at these same on-shell values and are therefore independent of the corresponding three-momentum magnitudes $p$ and $p'$. Thus, the off-shell dependence of the kernel is retained through the three-momenta, while the energy dependence is fixed by $\sqrt{s}$. The resulting ultraviolet (UV) divergence is regularized by multiplying the strong kernel by the Gaussian form factor
\begin{eqnarray}
F(p',p)=\exp\left(
-\frac{(p'^2-k'^2)+(p^2-k^2)}{\Lambda^2}
\right),
\end{eqnarray}
where $k,(k')$ denotes the incoming (outgoing) on-shell momentum. We take $\Lambda=790_{-28}^{+32}$ MeV, which provides a reasonable description of the on-shell amplitudes reported in Ref.~\cite{Albaladejo:2016lbb}.

\section{Coulomb interaction}
\label{sec:coulombinteraction}

For charged channels, the Coulomb interaction is incorporated following the momentum-space method of Ref.~\cite{Encarnacion:2026iur}, based on the Vincent--Phatak prescription \cite{Vincent:1974zz}. The long-range Coulomb potential is regularized using the auxiliary potential
\begin{equation}
V_{\rm short}(r)=V_S(r)+g(r)V_C(r),
\end{equation}
where we employ the Woods--Saxon regulator
\begin{equation}
g(r)=\frac{1+\exp(-R_{\rm WS}/b)}
{1+\exp[(r-R_{\rm WS})/b]},
\end{equation}
and throughout this work we set $R_{\rm WS}=7$ fm, $b=0.2$ fm and match the wave functions at $R_M=6$ fm, where the strong interaction is negligible. The corresponding $S$-wave momentum-space Coulomb potential is
\begin{equation}
    \!\!\! V_C^{\ell=0}(p',p) \! = \! \frac{4\pi Z_1 Z_2 \alpha}{pp'} \!\!\! \int_0^{\!\infty} \!\! \frac{dr}{r} g(r) \sin(pr) \sin(p'r)\,.
\end{equation}
On the mass shell, this potential exhibits a left-hand-cut (LHC) for $\kappa \geqslant 1/(2b) \simeq 500\,\text{MeV}$, with $k=i\kappa$. Therefore the WS-VP method is applicable only for $k^2$ above the branch point. For closed channels whose momenta lie inside the LHC, the Coulomb interaction can be neglected, since Coulomb effects become negligible far away from threshold.

Within the relativistic formulation adopted here, the Coulomb interaction enters the BSE through
\begin{equation} \label{eq:potentialVS+VC}
    V_{ij}(p',p)
    = V_{S,ij}^{\ell=0}(p',p)
    + \delta_{ij}
    \frac{\xi_i(p';s)\xi_i(p;s)}{\xi_i(k;s)}
    V_C^{\ell=0}(p',p) \, .
\end{equation}
where the factors
\begin{eqnarray}
    \xi_i(q;s) &=& 2E_{s,i} \frac{2\omega_{1i}(q)\omega_{2i}(q)}{\omega_{1i}(q)+\omega_{2i}(q)} \frac{s-(\omega_{1i}(q)+\omega_{2i}(q))^2}{k^2-q^2} \nonumber \\
    \xi_i(k;s) &=& \lim_{q\to k} \xi_i(q;s)=4 E_{s,i}\sqrt{s}
\end{eqnarray}
ensure consistency with the relativistic two-body equation (their derivation is given in the Appendix).

The solution of the regularized problem with the auxiliary potential
$V_{\rm short}$, denoted by $u_0^{\rm short}$, coincides with the exact
strong-plus-Coulomb wave function in the interaction region, up to an overall
normalization. For a single-channel system, this normalization is fixed by matching $u_0^{\rm short}$ to
the asymptotic strong-plus-Coulomb solution at $R_M$,
\begin{eqnarray}
    [u_0^{\rm asy}(k,r)]^* \!=\! e^{i\sigma_0} \Big[ \frac{F_0(\eta,kr)}{k} \!+\! f_{SC}(k) C_\eta^2 H_0^+(\eta,kr)\Big] \,,
\end{eqnarray}
where $F_0$, $G_0$ and $H_0^+=G_0+iF_0$ are the Coulomb functions, $\sigma_0=[\ln\Gamma(1+i\eta)-\ln\Gamma(1-i\eta)]/(2i)$ and $C_\eta^2=2\pi\eta[e^{2\pi\eta}-1]^{-1}$ are the Coulomb phase shift and the Gamow--Sommerfeld factor, respectively, and $f_{SC}$ is the Coulomb-modified strong scattering amplitude. Following Ref.~\cite{Encarnacion:2026iur}, the matching conditions yield
\begin{subequations}
\begin{align}
    A(k) &= \displaystyle\frac{ e^{i\sigma_0} }{
    \big[ H_0^+(\eta,kr), u_0^{\rm short}(k,r)^* \big]_{R_{\rm M}}}\,,
    \label{eq:coulombscreenedwavefunctionnormalization} \\
    f_{SC}^{-1}(k) &= -C_\eta^2 k\left(i +\frac{[G_0(\eta,kr),u_0^{\rm short}(k,r)^*]_{R_{\rm M}}}{[F_0(\eta,kr),u_0^{\rm short}(k,r)^*]_{R_{\rm M}}}\right)\ ,
    \label{eq:coulombscreenedwavefunctionfsc}
\end{align}
\end{subequations}
with $[\phi_1(r),\phi_2(r)]_{R_M}=\phi_1({R_M})\phi'_2({R_M})-\phi'_1({R_M})\phi_2({R_M})$ is the Wronskian evaluated at the matching radius.

The extension to coupled channels is completely analogous. We define $\mathcal U_0^{\rm short\dagger}$ as the matrix of coupled-channel solutions of the BSE, whose $ij$ element, $u^{\rm short}_{0,ij}(k,r)^* = r\,\psi_0^{i\leftarrow j}(k,r)^*$, is obtained from Eqs.~(\ref{eq:bethesalpeterequationwavefunction},\ref{eq:potentialVS+VC}). The complete wave-function matrix, at CM energy $\sqrt{s}$, is given by:
\begin{equation}
\!\!\! [\mathcal U_0(\sqrt s,r)]^\dagger\!=\! \left\{
\begin{array}{@{}ll@{}}
\mathcal A(\sqrt s)\,[\mathcal U^{\rm short}_0(\sqrt s,r)]^\dagger\,, & r \! \le \! R_{\rm M},\\[1.5ex]
\begin{aligned}[b]
e^{i\Sigma_0}\Big[ & K^{-1} \mathcal F_0(\sqrt s, r) \\
& + \mathcal B \, \mathcal H_0^+(\sqrt s,r)\Big]\,,
\end{aligned} & r \! > \! R_{\rm M}\,,
\end{array} \right.
\end{equation}
where $\mathcal B_{ij} = \mathcal K_{ij}C_{\eta_i}C_{\eta_j}f_{SC,ij}$. Here $\mathcal F_0$, $\mathcal H_0^+$, $K$, $C_\eta$ and $e^{i\Sigma_0}$ are diagonal matrices containing the corresponding Coulomb functions (with $H_0^+=G_0+iF_0$), momenta and Coulomb factors. Matching at $r=R_M$ determines the normalization matrix $\mathcal A$ and the Coulomb-modified scattering amplitudes $f_{SC}$,
\begin{subequations}\label{eq:phatakcoupledchannels}
\begin{align}
    \mathcal A & 
    = 
    -e^{i\Sigma_0} 
    \bigg[  
    \mathcal U^{\rm short\,\dagger}_0 \frac{d\mathcal H_0^+}{dr} 
    - 
    \frac{d\mathcal U^{\rm short\,\dagger}_0}{dr} \mathcal H_0^+ 
    \bigg]^{-1} \,,\\
    f_{SC}^{-1} & 
    = 
    - \mathcal S C_\eta 
    \Bigg\{ 
    i +  
    \bigg[ 
    \mathcal G_0[\mathcal U^{\rm short\,\dagger}_0]^{-1}\frac{d\mathcal U^{\rm short\,\dagger}_0}{dr} 
    - 
    \frac{d\mathcal G_0}{dr} 
    \bigg] \nonumber \\
    &\bigg[  
    \mathcal F_0[\mathcal U^{\rm short\,\dagger}_0]^{-1}\frac{d\mathcal  U^{\rm short\,\dagger}_0}{dr} 
    - 
    \frac{d\mathcal F_0}{dr} 
    \bigg]^{-1} 
    \Bigg\} 
    K C_\eta \mathcal S^{-1}\,,
\end{align}
\end{subequations}
where $\mathcal S = \mathrm{diag}\{\sqrt{\omega_{i1}\omega_{i2}}\}$. Here, all matrices are evaluated at $r=R_M$. Note that, as in the single channel case, $f_{SC}$ automatically satisfies unitarity,\footnote{Note that $[\mathcal U_0^{\rm short\,\dagger}]^{-1} d\mathcal U_0^{\rm short\,\dagger}/dr$ is a real matrix. Indeed, each row of $\mathcal U_0^{\rm short\,\dagger}$ can be identified with a solution vector of the coupled-channel Schr\"odinger equation and thus carries a common, $r$-independent overall phase. Consequently, $\mathcal U_0^{\rm short\,\dagger}$ can be written as $\mathcal U_0^{\rm short\,\dagger}=\mathcal E\mathcal R$, where $\mathcal E$ is an $r$-independent diagonal matrix containing these phases and $\mathcal R$ is a real matrix. Therefore, $[\mathcal U_0^{\rm short\,\dagger}]^{-1}d\mathcal U_0^{\rm short\,\dagger}/dr =\mathcal R^{-1}d\mathcal R/dr$,
which is manifestly real.} since the imaginary part of $f_{SC}^{-1}$ is the diagonal matrix with elements $-C_{\eta_i}^2k_i$. For closed channels, Eq.~\eqref{eq:phatakcoupledchannels} must be understood through the analytic continuation of $F_0$, $G_0$, $H_0^+$, $C_\eta$ and $\sigma_0$ at imaginary momenta.

\section{Results}\label{sec:results}

In this section, we first present the CFs obtained within the theoretical framework described above. We then analyze the contributions from the $I=1/2$ and $I=3/2$ isospin channels to the $D\pi$ scattering parameters and extend the analysis to the axial $D^*\pi$ sector using HQSS.
\begin{figure}[t]
    \centering
    \includegraphics[width=\linewidth]{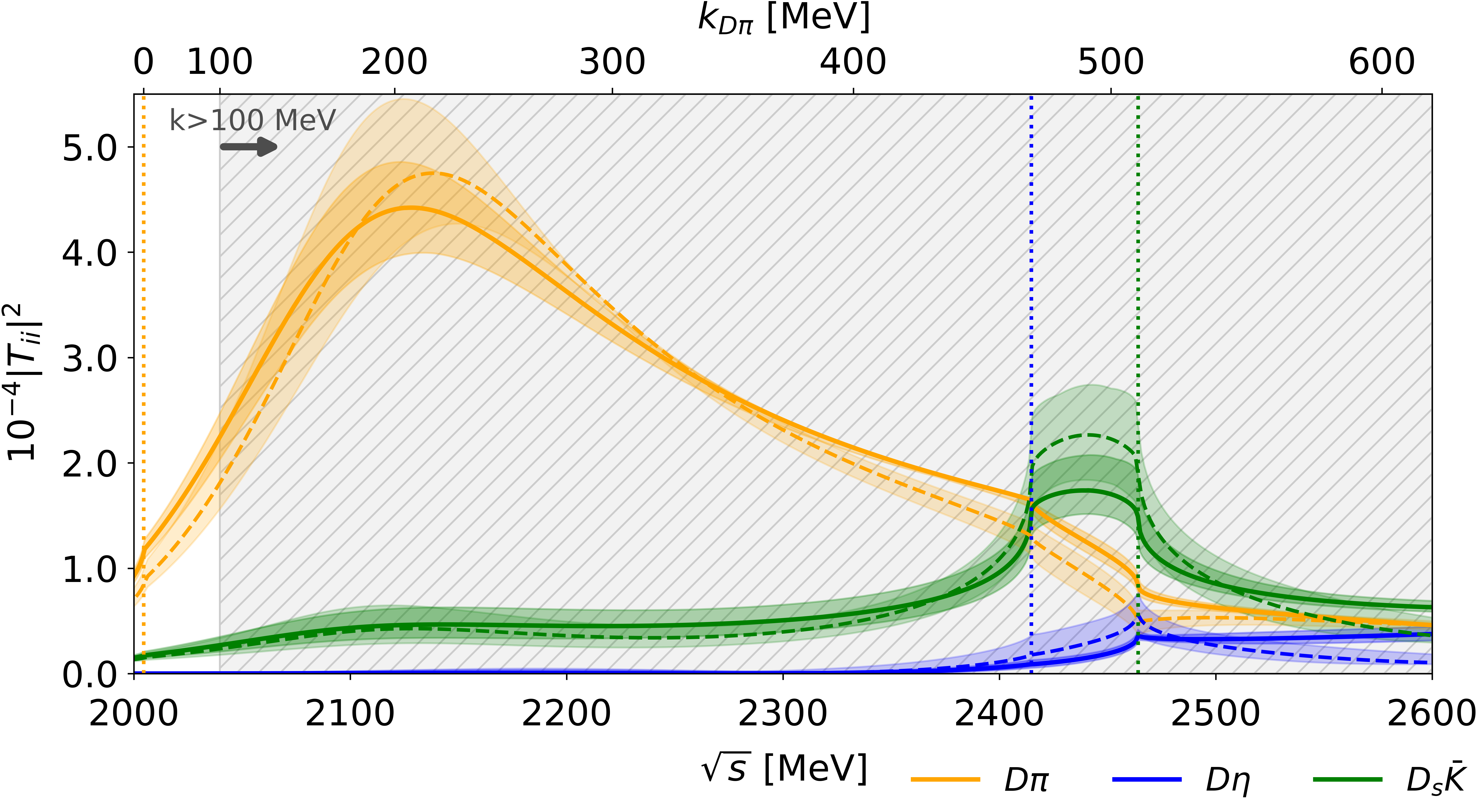}
    \caption{$T$-matrix modulus squared for the $I=1/2$ elastic transitions $D\pi$, $D\eta$, and $D_s\olsi{K}$. Solid lines correspond to the on-shell amplitudes obtained within our off-shell BSE scheme, while dashed lines show the results reported in Ref.~\cite{Albaladejo:2016lbb}. The error bands represent the 68\% confidence intervals obtained by propagating the uncertainties of the parameters in Table~\ref{tab:LECSfengkun}, together with the uncertainty in the UV cutoff $\Lambda$. The shaded region indicates the energy range for which the center-of-mass momentum of the $D\pi$ pair exceeds $100$ MeV.}
    \label{fig:tmatrix}
\end{figure}

In Fig.~\ref{fig:tmatrix}, we show the diagonal elements of the on-shell
$T$-matrix for the isospin $I=1/2$ channels, obtained with our off-shell
scheme (solid lines), together with those reported in
Ref.~\cite{Albaladejo:2016lbb} (dashed lines). Despite some differences,
the two results exhibit good overall agreement in both shape and magnitude.
This agreement is an essential consistency requirement for an off-shell
extension of the interaction, since the on-shell amplitudes are tightly
constrained by LQCD data. In particular, our model reproduces the
two-pole structure of the $D^*_0(2300)$ found in Ref.~\cite{Albaladejo:2016lbb}. We find two poles in the amplitude
on the second Riemann sheet, defined by \cite{Nieves:2001wt}
\begin{equation}
[T_{II}(s)]^{-1}
=
[T_I(s)]^{-1}
-
i\frac{k(s)}{4\pi\sqrt{s}} \,.
\end{equation}
in agreement with the double-pole pattern of the $D_0^*(2300)$ robustly established in Refs.~\cite{Albaladejo:2016lbb,Du:2017zvv}. Their positions in the complex plane are reported in Table~\ref{tab:polepositions}.
\begin{table}[hbt!]
    \centering
    \begin{tabular}{lll} \hline \hline
        \multicolumn{1}{c}{$D_0^*(2300)$ Poles} & \multicolumn{1}{c}{Lower pole} & \multicolumn{1}{c}{Higher pole} \\ \hline \\ [-3mm]
        This work & 
        $2083^{+3}_{-4}-111^{+7}_{-7}i$ & $2439^{+107}_{-55}-238^{+21}_{-21}i$ \\ [1mm]
        Albaladejo \textit{et al.} \cite{Albaladejo:2016lbb} & 
        $2105^{+6}_{-8}-102^{+10}_{-12}i$ & $2451^{+36}_{-26}-134^{+7}_{-8}i$\\ [1mm]
        Torres-Rincón \textit{et al.} \cite{Torres-Rincon:2023qll} & 
        $2092.4 - 129.5i$ & $2467.2 - 264.8i$\\ \hline \hline
    \end{tabular}
    \caption{Pole positions [MeV] $\sqrt{s}=M-i\Gamma/2$, on the second Riemann sheet obtained within our off-shell BSE scheme. The quoted uncertainties are obtained by propagating the uncertainties of the parameters in Table~\ref{tab:LECSfengkun}, together with the uncertainty in the UV cutoff $\Lambda$. The pole positions reported in Refs.~\cite{Albaladejo:2016lbb,Torres-Rincon:2023qll} are shown for comparison. }
    \label{tab:polepositions}
\end{table}

We also extract the scattering parameters from the amplitude $T_{SC}$, both with and without Coulomb effects. The elastic amplitude in channel $i$ is parametrized as
\begin{equation}
T_{SC,ii} = \frac{-8\pi\sqrt{s}\, C_{\eta_i}^2 e^{2i\sigma_{0,i}}}{-\frac{1}{a_{0,i}}+\frac{1}{2}r_{0,i}k_i^2 + \ldots - iC_{\eta_i}^2k_i -2\eta_i k_ih^\lambda(\eta_i)} \, ,
\end{equation}
 and $h^\lambda(\eta) = \sum_{n=1}^\infty \frac{\eta^2}{n(n^2+\eta^2)} - \ln(\lambda\eta) - \gamma_E$,
with $\lambda\equiv \mathrm{sgn}[Z_1Z_2]$, and $\gamma_E\simeq0.577$ is the Euler--Mascheroni constant.

We first examine the contributions of the $I=1/2$ and $I=3/2$ strong amplitudes to the elastic physical channels, finding:
\begin{subequations}
\begin{align}
    & f_{D^0\pi^0} = \frac{1}{3}f_{D\pi(I=1/2)} + \frac{2}{3}f_{D\pi(I=3/2)} \, ,\\
    & f_{D^+\pi^-} = \frac{2}{3}f_{D\pi(I=1/2)} + \frac{1}{3}f_{D\pi(I=3/2)} \, , \\
    & f_{D^+\pi^+} = f_{D\pi(I=3/2)} \, .
\end{align}
\end{subequations}
A substantial cancellation is expected in the linear combination $(a_{1/2}+2a_{3/2})$ of the $D\pi$ scattering lengths, as dictated by the low-energy chiral theorem for pion scattering off any particle other than a pion~\cite{Weinberg:1966kf}. Table~\ref{tab:scatteringparametersisospin} summarizes the scattering lengths in the isospin basis, together with the corresponding weighted combinations entering the physical channels, for three different levels of approximation: the Born approximation at LO, $T_{ij}=V_{ij}^S$, retaining only the Weinberg--Tomozawa $(s-u)$ contribution in Eq.~\eqref{eq:NLOstrongpotential}; the Born approximation including the NLO correction; and the full unitarized amplitude.
\begin{table}[hbt!]
    \centering
    \begin{tabular}{cccc} \hline\hline
     & Born (LO) & Born (LO+NLO) & Unitarized \\ \hline \\[-2.5mm]
     $a_0^{I=1/2}$ &
     $-0.236$ & $-0.233$ & $-0.423$ \\[1mm]
     $a_0^{I=3/2}$ &
     $0.118$ & $0.121$ & $0.105$ \\[1mm]
     $\frac{1}{3}a_0^{I=1/2} + \frac{2}{3}a_0^{I=3/2}$ &
     $0$ & $0.003$ & $-0.071$ \\[1mm]
     $\frac{2}{3}a_0^{I=1/2} + \frac{1}{3}a_0^{I=3/2}$ &
     $-0.118$ & $-0.115$ & $-0.247$ \\[1mm]
    \hline\hline
    \end{tabular}
    \caption{This table shows the strong scattering lengths [fm] in the isospin basis, $a_0^{I=1/2}$ and $a_0^{I=3/2}$, together with the corresponding weighted combinations for the physical $D^0\pi^0$ and $D^+\pi^-$ channels, obtained in the Born approximation with the LO interaction, the Born approximation including the NLO contribution, and the full unitarized amplitude. The unitarized results are in good agreement with those reported in the seminal work of Ref.~\cite{Liu:2012zya}.}
    \label{tab:scatteringparametersisospin}
\end{table}

At LO, the scattering length exhibits an exact cancellation in the isospin combination corresponding to the $D^0\pi^0$ channel, whereas the cancellation is only partial for the $D^+\pi^-$ channel. The inclusion of NLO terms slightly breaks the exact cancellation in the $D^0\pi^0$ channel, while unitarization further enhances the deviation from the LO result. As a consequence, the $D^0\pi^0$ interaction is strongly suppressed, whereas the $D^+\pi^-$ interaction, which is the focus of this work, retains most of its attractive character.
\begin{figure}
    \centering
    \includegraphics[width=\linewidth]{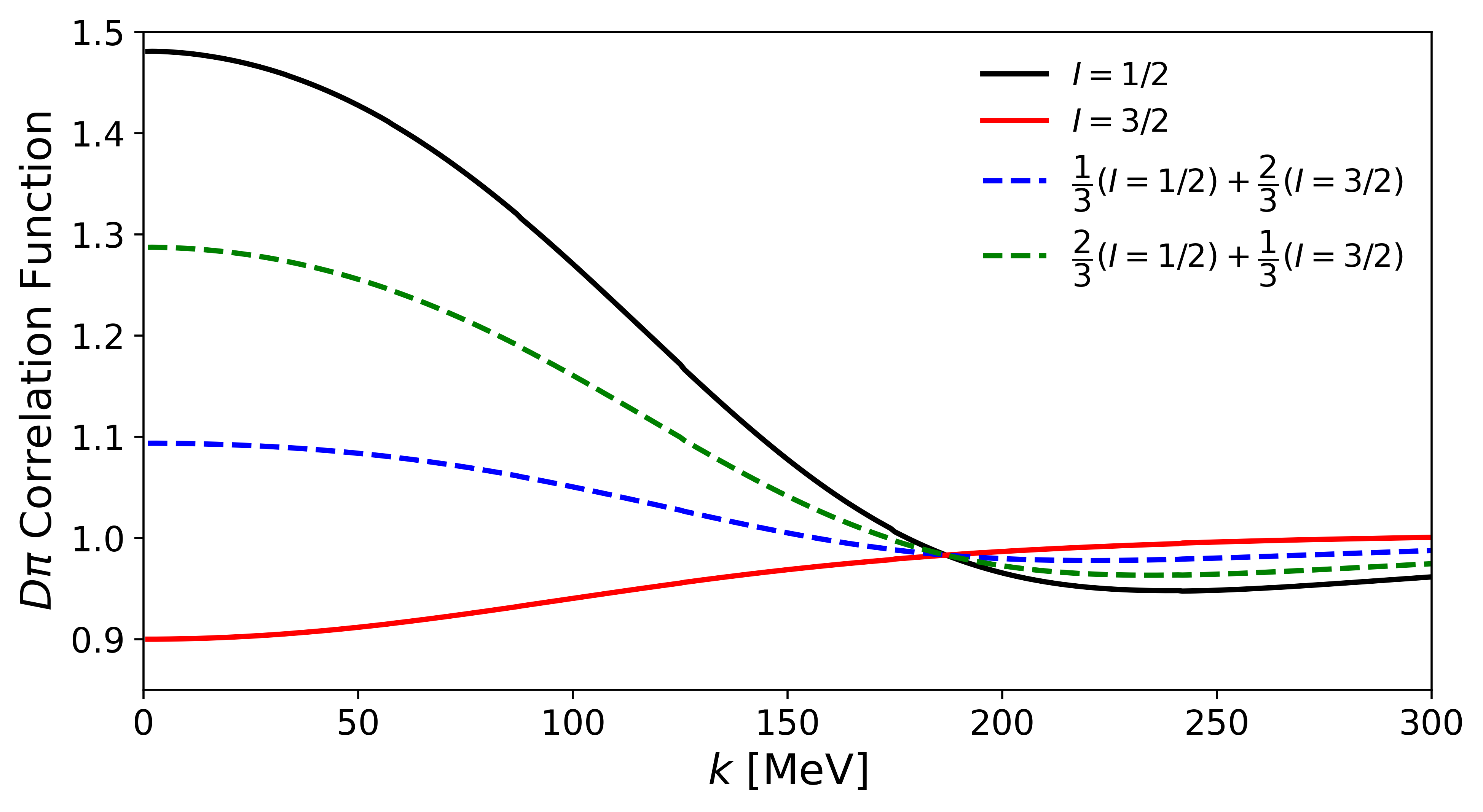}
    \caption{CFs for  physical $D\pi$ channels in the absence of Coulomb effects, showing the individual $I=1/2$ and $I=3/2$ contributions and the corresponding physical combinations, $C_{D^0\pi^0}=\frac{1}{3}C_{I=1/2}+\frac{2}{3}C_{I=3/2}$ and $C_{D^+\pi^-}=\frac{2}{3}C_{I=1/2}+\frac{1}{3}C_{I=3/2}$. The strong cancellation between the isospin amplitudes in the $D^0\pi^0$ combination leads to a strongly suppressed correlation signal. In this figure, a single-Gaussian source of $R=1$ fm has been used.}
    \label{fig:CFisospin}
\end{figure}

The impact of the isospin decomposition on the physical CFs is illustrated in Fig.~\ref{fig:CFisospin}, where the individual $I=1/2$ and $I=3/2$ contributions are shown together with the corresponding weighted combinations for the $D^0\pi^0$ and $D^+\pi^-$ channels. Consistent with the scattering lengths discussed above, the two isospin contributions largely cancel in the combination $\frac{1}{3}C_{I=1/2}+\frac{2}{3}C_{I=3/2}$ corresponding to $D^0\pi^0$, so that its CF remains close to unity. A similar cancellation between the attractive and repulsive isospin channels was found in the theoretical femtoscopy analysis of the $K^*_0(700)$ in Ref.~\cite{Albaladejo:2025lhn}, which was applied to the ALICE $K\pi$ data of Ref.~\cite{ALICE:2023eyl}. In contrast, the $D^+\pi^-$ combination, $\frac{2}{3}C_{I=1/2}+\frac{1}{3}C_{I=3/2}$, retains a sizable contribution from the attractive $I=1/2$ interaction, so its CF exhibits a substantially larger deviation from unity, making this channel more sensitive to the underlying dynamics of the lower $D^*_0(2300)$ pole. In the momentum range shown in the figure, the contribution of the higher channels $D_s\bar K$ and $D\eta$ to the $D\pi$ CFs is extremely small. Finally, our results for the $I=1/2$ and $I=3/2$ CFs are reasonably similar to those obtained in Ref.~\cite{Albaladejo:2023pzq}, obtained using on-shell $D\pi$, $D_s\bar{K}$, and $D\eta$ amplitudes of Ref.~\cite{Albaladejo:2016lbb} within the Lednicky–Lyuboshits approximation~\cite{Lednicky:1981su} supplemented by a UV cutoff.

We now examine the physical $D^+\pi^-$ and $D^+\pi^+$ channels including Coulomb effects. The resulting scattering parameters are shown in Table~\ref{tab:scatteringparameters} for the full strong-plus-Coulomb interaction and for the strong interaction alone. The latter scattering lengths are consistent with the isospin-limit values given in Table~\ref{tab:scatteringparametersisospin}, with the differences arising from the use of physical rather than isospin-averaged particle masses.
\begin{table}[hbt!]
    \centering
    \begin{tabular}{ccc} \hline\hline
         & $a_0$ [fm] & $r_0$ [fm] \\ \hline \\ [-2mm]
        $D^+\pi^-$ &
        $-0.255^{+0.012}_{-0.012} \!-\! 0.017^{+0.001}_{-0.001}i$ &
        $-14.5^{+0.6}_{-0.6} \!-\! 3.86^{+0.16}_{-0.16}i$ \\  [1.5mm]
        $D^+\pi^-$ ($\alpha\!=\!0$) &
        $-0.257^{+0.012}_{-0.012} \!-\! 0.017^{+0.001}_{-0.001}i$ &
        $-13.1^{+0.6}_{-0.6} \!-\! 3.85^{+0.16}_{-0.16}i$ \\ [1.5mm] \hline \\ [-1.7mm]
        $D^+\pi^+$  &
        $0.106^{+0.005}_{-0.005}$ &
        $21.8^{+1.0}_{-0.9}$ \\ [1.5mm]
        $D^+\pi^+$ ($\alpha\!=\!0$) &
        $0.107^{+0.005}_{-0.005}$ &
        $18.4^{+0.8}_{-0.7}$ \\ [1.5mm] 
        \hline\hline
    \end{tabular}
    \caption{Scattering parameters for the physical $D^+\pi^-$ and $D^+\pi^+$ systems with Coulomb interaction and without Coulomb interaction ($\alpha=0$). The quoted uncertainties are obtained by propagating the uncertainties of the parameters of the off-shell BSE scheme. }
    \label{tab:scatteringparameters}
\end{table}

\begin{figure*}
    \centering
    \includegraphics[width=\linewidth]{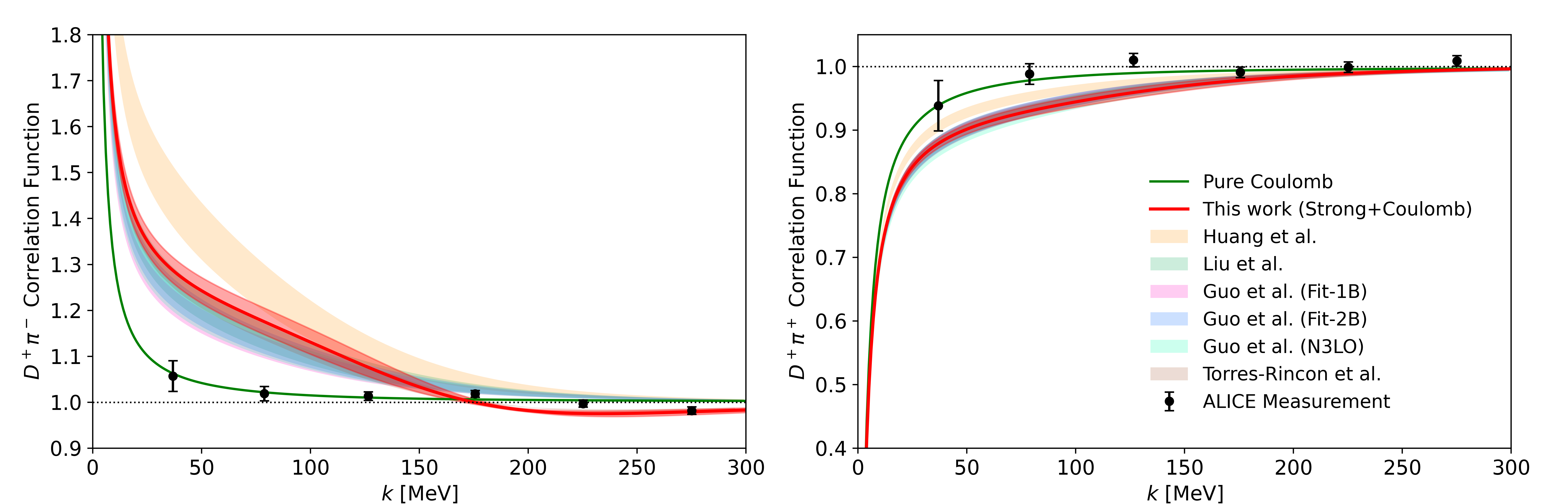}
    \caption{Our calculation of the $D^+\pi^-$ (left) and $D^+\pi^+$ (right) CFs is shown in red. The error bands represent the 68\% confidence intervals obtained by independently propagating the uncertainties associated with the theoretical model (the parameters in Table~\ref{tab:LECSfengkun} and $\Lambda$) and those of the source parameters ($x$, $R_1$, and $R_2$). The two contributions are then added in quadrature to obtain the displayed uncertainty bands. The experimental points are taken from Ref.~\cite{ALICE:2024bhk}, with the statistical and systematic uncertainties added in quadrature. The additional bands show the predictions of different theoretical approaches, taken from Fig.~4 of Ref.~\cite{ALICE:2024bhk} and based on Refs.~\cite{Huang:2021fdt,Liu:2011mi,Guo:2018tjx,Guo:2018kno,Torres-Rincon:2023qll}.}
    \label{fig:DPi_CF}
\end{figure*}

In Fig.~\ref{fig:DPi_CF} we show the $D^+\pi^-$ and $D^+\pi^+$ CFs (red curves and bands). For a consistent comparison with the experimental analysis, we use the source function reported in Refs.~\cite{ALICE:2024bhk,Chizzali:2025jog}, parametrized as a double Gaussian,
\vspace{-3pt}\begin{subequations}\begin{align}
S(r)     & = x \, S_G(r;R_1) + (1-x) \, S_G(r;R_2)\,, \\
S_G(r;R) & = (4\pi R^2)^{-3/2}\exp(-r/4R^2)\,,
\end{align}\end{subequations}
with $x=0.66^{+0.03}_{-0.02}$, $R_1=0.97^{+0.09}_{-0.08}$  fm, and $R_2=2.52^{+0.36}_{-0.20}$ fm. We also set the production weights entering Eq.~\eqref{eq:swaveKP} to $w_j=1$ for all channels. Although the production weights of the $D_s^+K^-$ and $D^0\eta$ channels could, in principle, be significantly smaller due to the lower production abundances expected for their larger masses relative to the elastic channel, we find that their contributions are negligible, amounting to less than $0.05\%$ of the total CF. The results are therefore insensitive to the precise values of these weights.

The central curves correspond to the best-fit values of the model parameters, while the uncertainty bands are obtained by independently propagating the uncertainties of the model parameters (LECs and $\Lambda$) and of the source parameters ($x$, $R_1$, and $R_2$). The two contributions are then added in quadrature. To provide a comprehensive illustration of the current situation, and inspired by the top panels of Fig.~4 in Ref.~\cite{ALICE:2024bhk}, we include the same theoretical curves for the $D^+\pi^+$ and $D^+\pi^-$ CFs, together with the genuine CF extracted in the ALICE analysis. As expected, our $D^+\pi^+$ and $D^+\pi^-$ CF predictions, based on a theoretically constrained $D\pi$ interaction (red curves), exhibit significant tension with the experimental data points, while remaining within the region spanned by the bulk of theoretical predictions. The origin of this discrepancy is investigated in the following subsection.

Finally, we extend the previous calculations to the $D^*\pi$ system. Within the HQSS framework adopted here, the S-wave $D^*\pi$ interaction in the $1^+$ channel is identical to the S-wave $D\pi$ interaction in the $0^+$ channel, up to replacing the $D$-meson masses by the corresponding $D^*$ masses. Consequently, the former develops a double-pole structure in the $2250$--$2550$ MeV region, analogous to that found for the $D_0^*(2300)$~\cite{Albaladejo:2016lbb}. The broad $D_1(2430)$, with $\Gamma\simeq 285$--$345$ MeV, listed by the PDG~\cite{ParticleDataGroup:2026mpi}, may therefore reflect the combined effect of these two nearby poles, in close analogy with the scalar case.\footnote{An additional, much narrower axial state, the $D_1(2420)$, with $\Gamma\sim 28-32$ MeV, is also listed by the PDG. The $D\rho$ channel, involving a light vector meson, could play an important role in its dynamics, as noted in Ref.~\cite{Albaladejo:2016lbb} (see also the recent works of Refs.~\cite{Khemchandani:2023xup,Brandao:2025cli}). This resonance is expected to contribute at a $D^*\pi$ CM momentum  $k\sim 350$ MeV. We only show our $D^*\pi$ CF results up to $k\leq 300$ MeV, well below the momentum region where the $D_1(2420)$ is expected to contribute. Thus, the omission of the $D\rho$ channel should not significantly affect the results presented here.} The resulting scattering parameters are shown in Table~\ref{tab:scatteringparametersDstar} for the full strong-plus-Coulomb interaction and for the strong interaction alone. The $D^{*+}\pi^-$ channel has a negative scattering length, reflecting the attractive character of the interaction, while the $D^{*+}\pi^+$ channel is characterized by a positive scattering length. As in the $D\pi$ case, the Coulomb interaction has only a small effect on the scattering lengths, while its impact on the effective range is more appreciable.
 
\begin{table}[hbt!]
    \centering
    \begin{tabular}{ccc} \hline\hline
         & $a_0$ [fm] & $r_0$ [fm] \\ \hline \\ [-2mm]
        $D^{*+}\pi^-$ &
        $-0.269 \!-\! 0.016i$ &
        $-14.5 \!-\! 3.90i$ \\  [1.5mm]
        $D^{*+}\pi^-$ ($\alpha\!=\!0$) &
        $-0.270 \!-\! 0.016i$ &
        $-13.1 \!-\! 3.88i$ \\ [1.5mm] \hline \\[-1.7mm]
        $D^{*+}\pi^+$  &
        $0.100$ &
        $21.9$ \\ [1.5mm]
        $D^{*+}\pi^+$ ($\alpha\!=\!0$) &
        $0.101$ &
        $18.4$ \\ [1.5mm]
        \hline\hline
    \end{tabular}
    \caption{Scattering parameters (central values) for the physical $D^{*+}\pi^-$ and $D^{*+}\pi^+$ systems, with and without the Coulomb interaction ($\alpha=0$).}
    \label{tab:scatteringparametersDstar}
\end{table}
The $D^{*+}\pi^+$ and $D^{*+}\pi^-$ CFs are shown in Fig.~\ref{fig:CFDstar}. We use the same double-Gaussian source function as in the $D\pi$ analysis, with parameters obtained from the experimental analysis of Ref.~\cite{ALICE:2024bhk}. The production weights are again set to $w_j=1$ for all channels. As before, a substantial deviation from the experimental data is observed, indicating a significant tension between the experimental analysis and the theoretical interaction.

Both the $D^*\pi$ scattering parameters and CFs show a clear equivalence with the corresponding $D\pi$ results, as expected from HQSS. The small differences arise mainly from the difference in reduced masses between the $D\pi$ and $D^*\pi$ pairs.
\begin{figure}
    \centering
    \includegraphics[width=\linewidth]{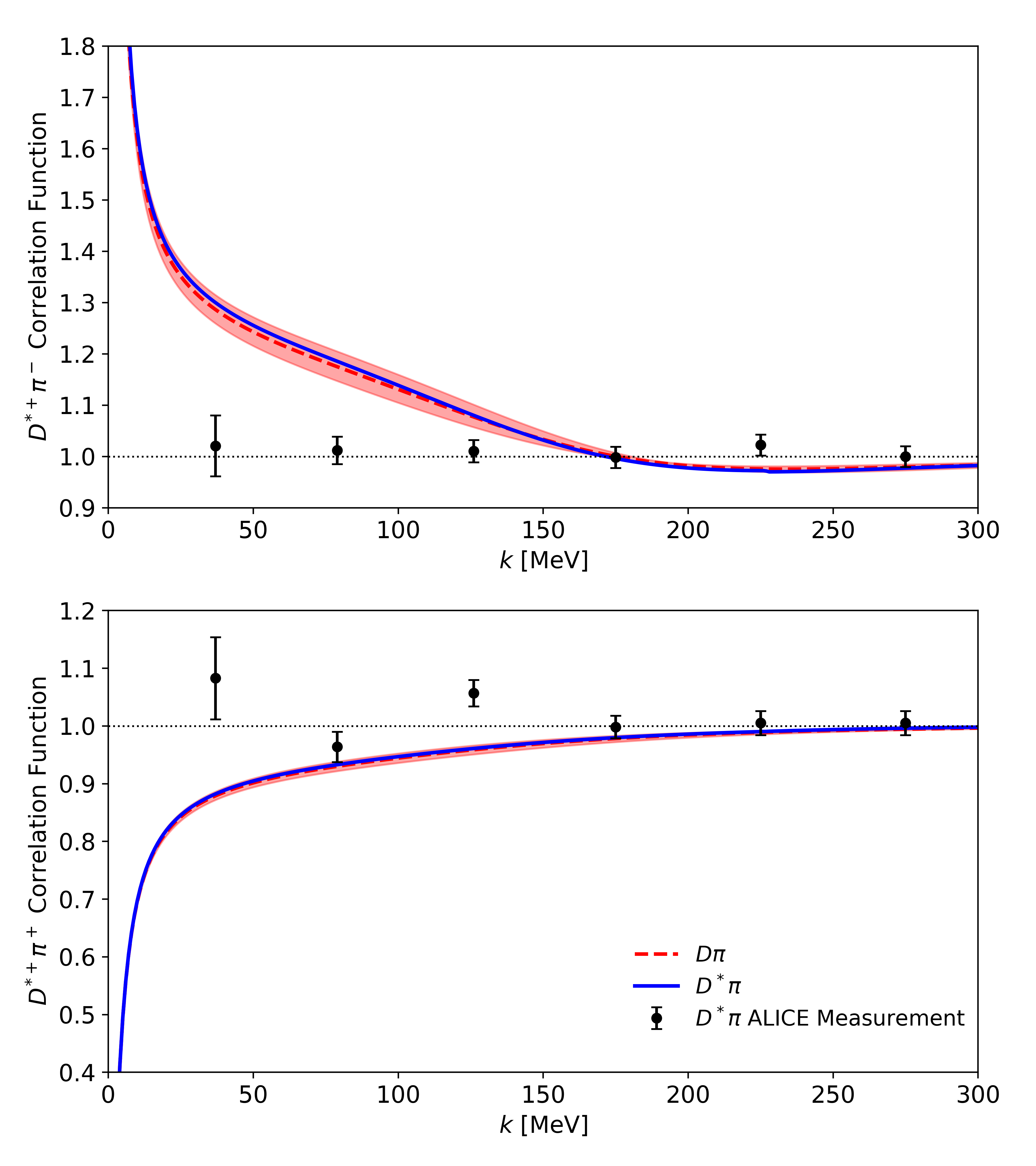}
    \caption{Our calculation of the $D^{*+}\pi^-$ (top) and $D^{*+}\pi^+$ (bottom) CFs is shown in blue. For comparison, the $D\pi$ results from Fig.~\ref{fig:DPi_CF} are shown in red, which are equivalent to the $D^*\pi$ CFs owing to HQSS. Experimental points are taken from Ref.~\cite{ALICE:2024bhk}, with statistical and systematic uncertainties added in quadrature.}
    \label{fig:CFDstar}
\end{figure}

\section{Comparison with experiment}\label{sec:results}

Our calculation is compatible with the other theoretical results shown in Fig.~\ref{fig:DPi_CF}.\footnote{This work differs from Ref.~\cite{Torres-Rincon:2023qll} primarily in the choice of LECs and UV form factor, and provides a better reproduction of the $D\pi$ on-shell amplitude of Ref.~\cite{Albaladejo:2016lbb}, particularly for $k>100$ MeV. It also incorporates an improved treatment of the Coulomb interaction and the kinematic factors $\mathcal K_{ij}$ in the inelastic wave functions. As shown in the figure, however, these changes have only a moderate impact on the predicted $D\pi$ CFs, well within the range of other theoretical uncertainties, and do not alleviate the pronounced tension with the ALICE measurements. Once the origin of this puzzle is understood, these differences may nevertheless prove relevant for a future precise extraction of the low-energy $D\pi$ scattering parameters.} However, none of them reproduces the experimental CFs. Since the measured CFs are found to be well described by the Coulomb interaction alone, Ref.~\cite{ALICE:2024bhk} interpreted this result as evidence that the residual strong interaction between charmed and light mesons is compatible with zero. Accordingly, the corresponding scattering lengths\footnote{Calculated as $a_0^{D^+\pi^+}=a_0^{I=3/2}$ and $a_0^{D^+\pi^-}=a_0^{I=3/2}/3+2a_0^{I=1/2}/3$ from the isospin-basis values reported in Ref.~\cite{ALICE:2024bhk}. The negative sign is introduced to match the sign convention with ours, that is, $f=-a_0$ at threshold.} were extracted as $a_0^{D^+\pi^-}=-0.017\pm0.021\pm0.007$ fm and $a_0^{D^+\pi^+}=-0.01\pm0.02\pm0.01\,\text{fm}$. These values are substantially smaller than those obtained within the unitarized NLO HMChPT calculation of Refs.~\cite{Liu:2012zya,Albaladejo:2016lbb}, which is consistent with the LQCD energy levels of Ref.~\cite{Moir:2016srx} and experimental data from LHCb~\cite{LHCb:2016lxy,Du:2017zvv}. The corresponding scattering lengths are calculated within our framework and compiled in Table~\ref{tab:scatteringparameters}. The same conclusions apply to the $D^*\pi$ system, as illustrated by the comparison of our theoretical predictions with the ALICE measurements in Fig.~\ref{fig:CFDstar}. For brevity, we focus below on the $D\pi$ system, with the results for $D^*\pi$ following analogously from  HQSS.

The detailed analysis of Ref.~\cite{Chizzali:2025jog} identifies a possible source of the discrepancy in a scenario where charmed mesons hadronize earlier than light mesons. In this picture, the $D$ mesons propagate before final-state interactions set in, effectively increasing the source size. As illustrated in Fig.~4.30 of Ref.~\cite{Chizzali:2025jog}, the average hadronization time difference between $\pi$ and $D$ mesons is $\Delta t\sim0.4$ fm/$c$. Even assuming propagation at the speed of light, this would increase the source radius by at most $\Delta R=\sqrt{R^2+(c\Delta t)^2}-R\lesssim0.1$ fm for typical source sizes. By contrast, compatibility with our results at the $1\sigma$ level requires a Gaussian source with an effective radius $R\gtrsim2.8$ fm. Starting from a nominal $R\simeq1$ fm, achieving such an increase with $\Delta t=0.4$ fm/$c$ would require a propagation velocity of $v\simeq6.5c$, which is clearly incompatible with causality.

However, the apparent tension may originate from an assumption in the experimental analysis which does not hold for the $D\pi$ system and potentially biases the extracted scattering length, as we explain next. In Ref.~\cite{ALICE:2024bhk}, the raw CF is modeled as a combination of several femtoscopic contributions: side-band effects from particle reconstruction, $\lambda_{\rm SB}C_{\rm SB}$; $D^*\to D$ feed-down, $\lambda_{D \leftarrow D^*}C_{D \leftarrow D^*}$; an additional flat background, $\lambda_{\rm flat}$; and the genuine $D\pi$ contribution, $\lambda_{\rm gen}C_{\rm gen}$. These are supplemented by a non-femtoscopic background, $C_{\rm NF}$, obtained from Monte Carlo (MC) simulations of the other processes contributing to the measured signal, $C_{\rm MC}$. The resulting parametrization is
\begin{align}
C_{\rm raw} & = \lambda_{\rm SB}C_{\rm SB}
+ C_{\rm NF}
\left(
\lambda_{\rm gen}C_{\rm gen}
+ \lambda_{D \leftarrow D^*}C_{D \leftarrow D^*}
+ \lambda_{\rm flat}
\right)\,, \nonumber \\
C_{\rm NF} & = N(C_{\rm MC}+ak^2)\,.\label{eq:Crawexperimental}
\end{align}
The parameters $N$ and $a$, which rescale and modify the MC-based correlation function, are fitted to the raw data under the assumption that $C_{\rm gen}=1$ for $k>k_{\rm min}$. While this approximation is generally justified once the interaction becomes negligible, the required momentum scale is system dependent. For $D\pi$, the choice adopted by ALICE in the analysis of Ref.~\cite{ALICE:2024bhk}, $k_{\text{min}}=100\,\text{MeV}$, is clearly not appropriate, as substantial strong-interaction effects persist well above this value. In fact, the peak associated with the lowest $D_0^*(2300)$ pole is located around $k\simeq 200\,\text{MeV}$. As shown in Fig.~\ref{fig:tmatrix}, setting $C_{\rm gen}=1$ for $k>100$ MeV therefore removes a significant part of the genuine strong-interaction signal, and, in particular, most of the strength of the lowest pole. Since this condition is subsequently used to constrain the background and extrapolate the genuine CF to lower momenta, the extracted CF can be likely biased toward a Coulomb-dominated behavior.\footnote{Furthermore, in the experimental analysis, the $D^{\ast} \to D$ feed-down is modeled assuming Coulomb interaction only, as the $D^{\ast}\pi$ CF measured by ALICE is well described by the corresponding Coulomb-only calculation. If the $D^{\ast}\pi$ interaction is as sizable as expected from HQSS and obtained in Sect.~\ref{sec:results} (see Fig.~\ref{fig:CFDstar}), this contribution would also deviate from the Coulomb-only estimate, potentially introducing an additional systematic effect in the extraction.}

From the theoretical perspective, constructing a realistic interaction satisfying $C_{\rm gen}=1$ for $k>100$ MeV would require suppressing the strong dynamics above this momentum scale. 
\begin{figure}
    \centering
    \includegraphics[width=\linewidth]{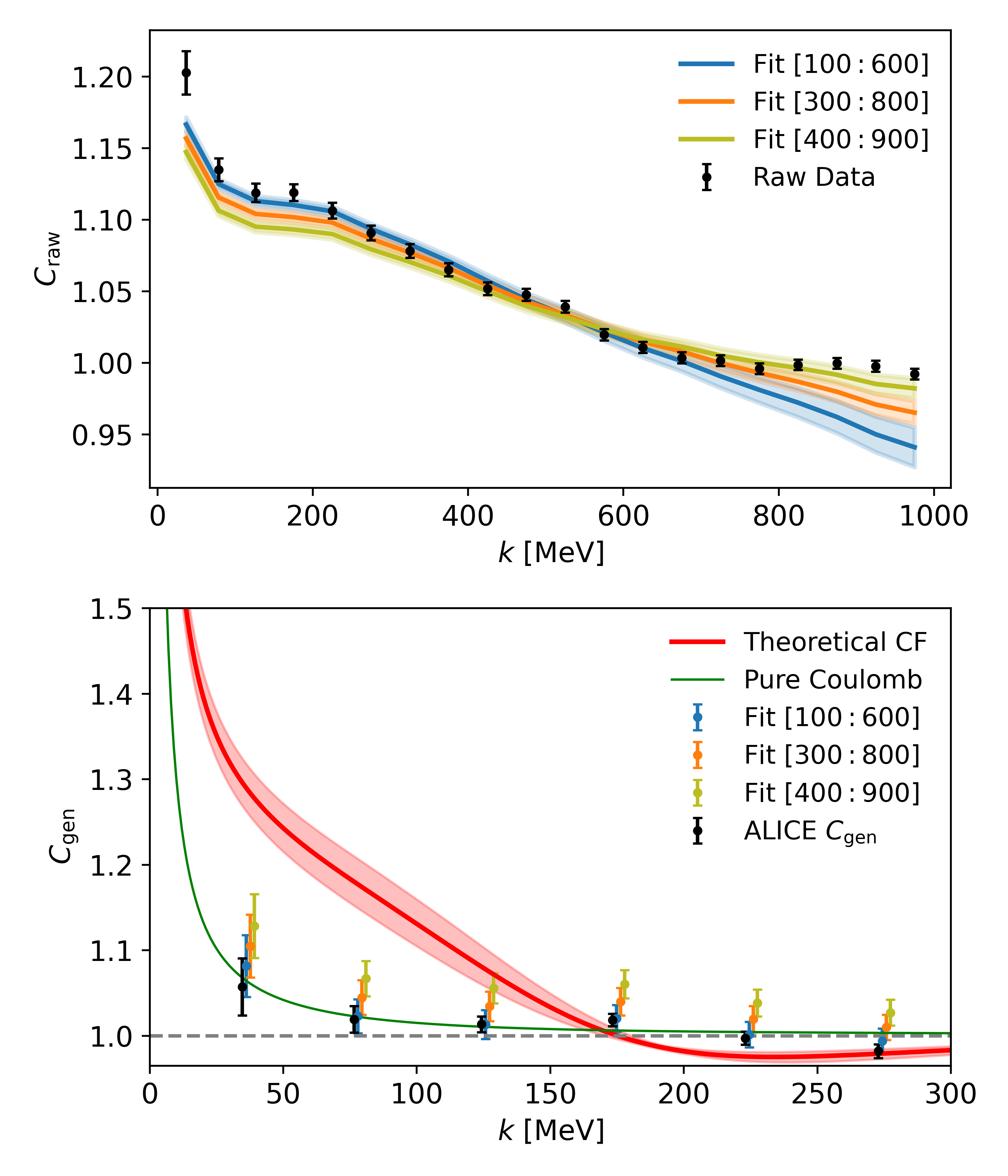}
    \caption{Fits to the raw data from Refs.~\cite{ALICE:2024bhk,Chizzali:2025jog} for different $k$ intervals, where $k$ denotes the $D\pi$ relative momentum. Upper panel: total background from Eq.~\eqref{eq:Crawexperimental} with $C_{\rm gen}=1$. Lower panel: corresponding extracted $C_{\rm gen}$, compared with the ALICE result~\cite{ALICE:2024bhk}, our theoretical result, and the pure-Coulomb case. In the lower panel, data points are horizontally offset for clarity but correspond to the same $k$ bins.}
    \label{fig:ALICEFit_Q0}
\end{figure}
A direct way to assess the impact of this assumption is to repeat the extraction procedure of Ref.~\cite{ALICE:2024bhk}, varying only the value of $k_{\rm min}$. We stress that this does not constitute a reanalysis of the experimental data, but rather a test of the robustness of the conclusions of Ref.~\cite{ALICE:2024bhk}. The resulting genuine CF should therefore not be interpreted as the physical correlation function. The ingredients entering Eq.~\eqref{eq:Crawexperimental} can be obtained from Refs.~\cite{ALICE:2024bhk, Chizzali:2025jog}: the $\lambda$ parameters are taken from Table~III of Ref.~\cite{ALICE:2024bhk}, while $C_{\rm raw}$, $C_{\rm SB}$, $C_{\rm MC}$, and $C_{D \leftarrow D^*}$ are extracted from Figs.~4.17, 4.19, 4.22, and 4.23 of \cite{Chizzali:2025jog}, respectively. We then repeat the fitting procedure using different momentum windows. While the original analysis uses $k\in[100,600]$ MeV, with variations of the upper limit to estimate the associated systematic uncertainty, we additionally consider the intervals $[300,800]$ MeV and $[400,900]$ MeV. The resulting fit parameters are reported in Table~\ref{table:fitNa}.
\begin{table}[t]
    \centering
    \begin{tabular}{ccc}\hline\hline
        $k$ [MeV]& $N$ & $a$ [MeV$^{-2}$] \\ \hline \\[-3mm]
        $[100,600]$ & $1.036\pm0.004$ & $(-1.28\pm0.19)\cdot 10^{-7}$ \\
        $[300,800]$ & $1.023\pm0.004$ & $(-8.14\pm1.07)\cdot 10^{-8}$ \\
        $[400,900]$ & $1.010\pm0.005$ & $(-4.45\pm0.90)\cdot 10^{-8}$ \\\hline\hline
    \end{tabular}
    \caption{Fitted parameters $a$ and $N$ in Eq.~\eqref{eq:Crawexperimental} for the different $k$ intervals considered.}
    \label{table:fitNa}
\end{table}

The results are shown in Fig.~\ref{fig:ALICEFit_Q0}. The choice of momentum interval over which $C_{\rm gen}$ is assumed to be unity directly affects the extracted total background (upper panel) and, consequently, the genuine CF obtained by extrapolation to lower momenta (lower panel). As a validation of our procedure, repeating the original fit (blue points and band) yields a genuine CF very similar to that reported in Ref.~\cite{ALICE:2024bhk}, consistent with both their result and a purely Coulomb interaction.

Interestingly, increasing the lower edge of the fit interval systematically raises the extracted genuine CF, progressively shifting it away from the pure-Coulomb scenario and towards the theoretical results. However, it does not bring them into complete agreement, as could be expected, since this exercise is not intended to provide a revised extraction of the genuine CF, but rather to assess the sensitivity of the experimental conclusions to the assumption $C_{\rm gen}=1$ above $k=100$ MeV. The pronounced dependence on the fit window demonstrates that the conclusion of a negligible strong interaction is not robust under this assumption. This issue calls for a new quantitative determination of the genuine CF through a dedicated reanalysis of the experimental data.

\section{Conclusions}

We revisit the $D\pi$ system from a femtoscopic perspective, motivated by the recent ALICE measurement indicating a negligible residual strong interaction, in apparent tension with theoretical predictions of a sizable $D\pi$ interaction. We first develop a relativistic momentum-space framework for femtoscopic CFs in coupled channels, incorporating both strong and Coulomb interactions. The formalism extends the Vincent--Phatak prescription to coupled channels within the three-dimensional BSE while preserving the correct Coulomb asymptotics.

We have applied the framework developed in this work to the $D\pi$ system. Given the pronounced tension with the ALICE measurement, a central question is whether this discrepancy could originate from the theoretical construction of the CF itself. This issue is particularly relevant because the underlying strong interaction is tightly constrained by chiral symmetry and LQCD and exhibits the well-established two-pole structure of the $D_0^*(2300)$. Our calculation solves the off-shell BSE and reproduces with high accuracy the on-shell scattering amplitudes of Ref.~\cite{Albaladejo:2016lbb}. It also provides a more complete treatment of the CF, incorporating the relativistic coupled-channel dynamics, Coulomb interaction, and transition wave functions consistently within the Bethe--Salpeter framework. We find, however, that these improvements have only a moderate impact on the predicted CFs. Within the current uncertainties of the strong-interaction model, our results remain consistent with previous theoretical calculations, while still showing a clear discrepancy with the genuine CF extracted by the ALICE Collaboration.

The persistence of this discrepancy indicates that it cannot not resolved by the specific theoretical ingredients examined here. Once this puzzle is resolved, however, the framework developed in this work may prove valuable for a future precision extraction of the low-energy $D\pi$ scattering parameters. We also stress that theoretical CFs can be affected by ambiguities in the treatment of the wave function at short distances. Within the Koonin--Pratt formalism, such ambiguities should be compensated by an appropriate choice of the source describing the emission of the interacting hadron pair~\cite{Epelbaum:2025aan, Molina:2025lzw, Albaladejo:2025kuv, Encarnacion:2026iur}. We nevertheless do not expect these effects to account for the pronounced tension with the ALICE measurement. We further extend the study to the $D^*\pi$ system, relying on HQSS to determine its scattering parameters and correlation functions, and find a similar tension with the ALICE measurements.

Finally, we show that this discrepancy, or at least a part of it, may be traced to the assumption made in Ref.\,\cite{ALICE:2024bhk} that $C_{\rm gen}=1$ for $k>100$ MeV. The lower $D_0^*(2300)$ pole still has a significant impact in this momentum region, so imposing this condition effectively removes part of the genuine strong-interaction signal and biases the extracted CF. We find that the extracted $D\pi$ signal depends significantly on the chosen fitting interval, demonstrating that the extraction procedure is sensitive to this assumption. This sensitivity indicates that a dedicated reanalysis of the experimental data is needed to establish  reliable $D\pi$ and $D^*\pi$ CFs and, consequently, robust scattering parameters. This issue is particularly relevant to the use of femtoscopy as a tool for constraining low-energy hadron interactions involving pions, whose dynamics is tightly constrained by chiral symmetry. In this context, the difficulties encountered in Ref.~\cite{Albaladejo:2025lhn} in describing the ALICE $K\pi$ CF measurements~\cite{ALICE:2023eyl} are especially relevant. Despite employing a state-of-the-art on-shell dispersive amplitude constrained by available $K\pi$ scattering data~\cite{Pelaez:2020gnd}, a satisfactory description of the measured CF could not be achieved.  Taken together, these observations underscore the need to clarify the origin of the tension between theoretical CFs and the ALICE measurements, further strengthening the prospects of femtoscopy as a reliable quantitative tool for precision studies of low-energy hadron interactions.

\begin{acknowledgments}
    \vspace{0.5cm}

This work is part of the Grants {\small PID2023-147458NB-C21} and {\small CEX2023-001292-S} funded by {\small MICIU/AEI/10.13039/501100011033} and by ERDF/EU, as well as of the Grant {\small CIPROM/2023/59} funded by Generalitat Valenciana {\small 10.13039/501100003359}. %
M.\,A.\,acknwoledges the \guillemotleft{}Ramón y Cajal\guillemotright{} program Grant {\small RYC2022-038524-I} funded by {\small MICIU/AEI/10.13039/501100011033} and by ESF+, and the \guillemotleft{}Atracción de Talento\guillemotright{} program Grant {\small PIE 20245AT019} funded by {\small CSIC 10.13039/501100003339}. %
M.\,A. and A.\,F. warmly thank the support from ACVJLI.
\end{acknowledgments}

\bibliographystyle{apsrev4-1_MOD}
\bibliography{RefsDPi}

\appendix
\section*{Appendix: Additional Details on the Calculation of the CFs}

\subsection{Non-relativistic limit of the wave functions}
\label{sec:kij}

For a given half-off-shell coupled-channel $T$-matrix, $T_{ij}(k,q; s)$, the $\ell=0$ wave function for the transition $j\to i$ is obtained from the Lippmann--Schwinger equation in the non-relativistic case and from the BSE in the relativistic case,
\begin{eqnarray} \label{eq:S1}
    [\psi_0^{i\leftarrow j}\!(k,r)]^*
    \!\! &=& \! j_0(kr)\delta_{ij} \nonumber \\
    &+& \!\! \int \!\!\!\frac{d^3q}{(2\pi)^3}\,
    T_{ij}(k,q; s) G_j(q) j_0(qr),
\end{eqnarray}
where the boundary condition is chosen such that channel $i$ is the observed channel, with its on-shell momentum given by $k\equiv k_i=\lambda^\frac12(s,m_{1i}^2,m_{2i}^2)/(2\sqrt{s})$  where $\lambda$ denotes the Källén function\footnote{In the non-relativistic limit $k_i\sim \sqrt{2\mu_i(\sqrt{s}-m_{1i}-m_{2i})}$}.
In Eq.~\eqref{eq:S1}, the two body propagator can be written as
\begin{subequations}
\begin{align}
    & \!\!\! G_j^{\rm LS}(q) \!=\! \frac{1}{m_{1i} \!+\! m_{2i} \!-\! m_{1j} \!-\! m_{2j} \!+\! k^2/2\mu_i \!-\! q^2/2\mu_j \!+\! i\epsilon} \,, \\
    & \!\!\! G_j^{\rm BS}(q) \!=\! \frac{\omega_{1j}(q) \!+\! \omega_{2j}(q)}{2\omega_{1j}(q)\omega_{2j}(q)} \frac{1}{s \!-\! (\omega_{1j}(q) \!+\! \omega_{2j}(q))^2 \!+\! i\epsilon},  \\
    & \!\!\!  \omega_{aj}(q)=\sqrt{m_{aj}^2+q^2},\, a=1,2 \nonumber
\end{align}
\end{subequations}
The asymptotic behavior of the wave function is determined by the pole of the propagator in Eq.~\eqref{eq:S1}, since the regular part vanishes (Riemann--Lebesgue lemma). For the LS propagator, the residue at $q=k_j$ is
\begin{equation}
\operatorname*{Res}_{q=k_j}
\left[q\,T^{\rm LS}_{ij}(k,q; s)G_j^{\rm LS}(q)\right]
=
-\mu_j\,T^{LS}_{ij}(k),
\end{equation}
with $T^{\mathrm{LS}}_{ij}(k)$ denoting the LS on-shell amplitude, leading to~\cite{Landau:1989}\footnote{Note that the textbook discusses the $i\to j$ transition, whereas here we are interested in the reverse transition, $j\to i$.}
\begin{eqnarray}\label{eq:S5}
[\psi_0^{i\leftarrow j}(k,r)]^*
\xrightarrow[r\to\infty]{} && \!\!\!\!
\delta_{ij}j_0(kr)
-\frac{\mu_j}{2\pi}
T^{LS}_{ij}(k)
\frac{e^{ik_jr}}{r} \nonumber \\
= && \!\!\!\! \delta_{ij}j_0(kr) \!+\! \sqrt{\frac{\mu_j}{\mu_i}} f_{ij}(k) \frac{e^{ik_jr}}{r}
\end{eqnarray}
where the final CM momentum $k_j$ is, in the non-relativistic limit, given by $k_j \sim \sqrt{2\mu_j(\sqrt{s}-m_{1j}-m_{2j})}$.

Similarly, for the BS propagator, evaluating the pole residue at $\sqrt{s}=\omega_{1j}(k_j)+\omega_{2j}(k_j)$, with $k_j=\lambda^{1/2}(s,m_{1j}^2,m_{2j}^2)/(2\sqrt{s})$, yields
\begin{equation}
\operatorname*{Res}_{q=k_j}
\left[q\,T^{BS}_{ij}(k,q;s)G_j^{\rm BS}(q)\right]
=
-\frac{T^{BS}_{ij}(k)}{4\sqrt{s}},
\end{equation}
so that
\begin{eqnarray}\label{eq:S7}
[\psi_0^{i\leftarrow j}(k,r)]^*
\xrightarrow[r\to\infty]{} && \!\!\!\!
\delta_{ij}j_0(kr)
-\frac{T^{BS}_{ij}(k)}
{8\pi\sqrt{s}}
\frac{e^{ik_jr}}{r} \nonumber \\
= && \!\!\!\! \delta_{ij}j_0(kr) + f_{ij}(k) \frac{e^{ik_jr}}{r}
\end{eqnarray}
In Eqs.~(\ref{eq:S5}, \ref{eq:S7}) we have introduced the relation between the on-shell quantum-mechanical scattering amplitude, $f$, and the LS and BS $T$-matrices:
\begin{eqnarray}
    f_{ij}(k) = - \frac{\sqrt{\mu_i\mu_j}}{2\pi} T_{ij}^{\rm LS}(k) = -\frac{T_{ij}^{\rm BS}(k)}{8\pi\sqrt{s}} \,.
 \end{eqnarray}
The factor $\sqrt{\mu_j/\mu_i}$ appearing in Eq.~(\ref{eq:S5}) originates from the reduced-mass dependence of the residue of the LS propagator, whereas the BS propagator residue depends only on $\sqrt{s}$. As a consequence, the asymptotic wave function in Eq.~(\ref{eq:S7}) does not reproduce the expected non-relativistic limit. This affects observables depending explicitly on the wave functions, such as the femtoscopic CF. To recover the correct non-relativistic limit, we introduce a kinematic factor $\mathcal K_{ij}$ into the definition of the inelastic wave functions within the BS scheme
\begin{eqnarray}
    \!\!\!\!\!\!\!\!\!\!\!
    [\psi_0^{i\leftarrow j}\!(k,\!r)]^* \!&=&\! j_0(kr)\delta_{ij} \nonumber \\
    &+& \! \mathcal K_{ij} \!\!\int \!\!\!\frac{d^3 q}{(2\pi)^3} T^{BS}_{ij}(k,q;s) G^{BS}_j\!(q) j_0(qr).
\end{eqnarray}
where the kinematic factor is defined as
\begin{eqnarray}
    \mathcal K_{ij} = \sqrt{\frac{\omega_{1j}(k_j)\,\omega_{2j}(k_j)}{\omega_{1i}(k)\,\omega_{2i}(k)}}
\end{eqnarray}
The introduction of this factor can be justified on several grounds. First, as discussed above, it ensures the correct non-relativistic limit for observables constructed from the wave functions. Second, it follows naturally from the single-particle Klein--Gordon equations: the associated relativistic probability current fixes the relative flux through the particle velocities, leading to the kinematic factor $\mathcal K_{ij}$, which plays the role of the relativistic counterpart of $\sqrt{\mu_j/\mu_i}$.\footnote{The kinematic factor $\mathcal K_{ij}$ is fixed by requiring consistency with the
standard relation between the differential cross section and the scattering
amplitude,
\begin{equation}
\frac{d\sigma_{j\leftarrow i}}{d\Omega}
=
\frac{k_j}{k_i}|f_{j\leftarrow i}|^2.
\label{eq:cross}
\end{equation}
We work in the final-state wave-function convention commonly adopted in
femtoscopy,
\begin{equation}
[\psi^{i\leftarrow j}]^*
\sim
e^{i\vec k_i\cdot\vec r}
+
\mathcal K_{ij} f_{i\leftarrow j}
\frac{e^{ik_j r}}{r},
\label{eq:psi}
\end{equation}
in which the plane wave observed in channel $i$ is propagated backward in
time toward the produced channel $j$ (see Ref.~\cite{Albaladejo:2024lam}).
The complex-conjugated wave function can therefore be interpreted as a
genuine scattering solution for the time-reversed process $j\leftarrow i$,
with channel $i$ acting as the incoming channel and channel $j$ as the
outgoing one. For this scattering solution, the differential cross section
follows from the ratio of outgoing to incoming flux,
\begin{equation}
\frac{d\sigma_{j\leftarrow i}}{d\Omega}
=
\frac{J_{\rm out}r^2}{J_{\rm in}},
\end{equation}
with $J_{\rm in}$ and $J_{\rm out}$ evaluated in channels $i$ and $j$,
respectively. From the asymptotic form of $[\psi^{i\leftarrow j}]^*$,
\begin{equation}
J_{\rm in}\propto v_i,
\qquad
J_{\rm out}\propto
\frac{v_j}{r^2}\mathcal K_{ij}^2
|f_{i\leftarrow j}|^2,
\end{equation}
where $v_{i(j)}$ denotes the relative two-hadron velocity in channel
$i(j)$. It follows that
\begin{equation}
\frac{d\sigma_{j\leftarrow i}}{d\Omega}
=
\frac{v_j}{v_i}
\mathcal K_{ij}^2
|f_{i\leftarrow j}|^2
=
\frac{v_j}{v_i}
\mathcal K_{ij}^2
|f_{j\leftarrow i}|^2,
\end{equation}
where the last equality follows from the reciprocity relation
$f_{i\leftarrow j}=f_{j\leftarrow i}$, which holds under time-reversal
invariance for amplitudes normalized as in Eq.~\eqref{eq:cross}.
Comparison with Eq.~\eqref{eq:cross} then fixes $\mathcal K_{ij}$ once the
relative velocity is expressed in terms of the channel momentum.

In the Schr\"odinger case, $\vec v=\vec k/m$, so that
\begin{equation}
v_{i(j)}
=
|\vec v_{1i(j)}-\vec v_{2i(j)}|
=
k_{i(j)}
\left(
\frac{1}{m_{1i(j)}}+
\frac{1}{m_{2i(j)}}
\right)
=
\frac{k_{i(j)}}{\mu_{i(j)}},
\end{equation}
and consistency with Eq.~\eqref{eq:cross} gives
\begin{equation}
\mathcal K_{ij}
=
\sqrt{\frac{\mu_j}{\mu_i}}.
\end{equation}

In the Klein--Gordon case, working in the center-of-mass frame, the velocity
of each particle is $\vec v_a=\vec k_a/\omega_a$. The corresponding relative
two-hadron speed is therefore
\begin{eqnarray}
v_{i(j)}
&=&
|\vec v_{1i(j)}-\vec v_{2i(j)}|
=
k_{i(j)}
\left(
\frac{1}{\omega_{1i(j)}}+
\frac{1}{\omega_{2i(j)}}
\right)
\nonumber\\
&=&
\frac{k_{i(j)}\sqrt{s}}
{\omega_{1i(j)}\omega_{2i(j)}},
\end{eqnarray}
where we have used $\omega_{1i}+\omega_{2i} =
\omega_{1j}+\omega_{2j} = \sqrt{s}.$ Consequently,
\begin{equation}
\mathcal K_{ij}
=
\sqrt{
\frac{\omega_{1j}\omega_{2j}}
{\omega_{1i}\omega_{2i}}
}.
\end{equation}} Finally, since the CF depends on $|\psi|^2$, the wave functions must be interpreted as probability amplitudes. Accordingly, the Klein--Gordon fields must carry the standard relativistic factors $\sqrt{2E}$ for the initial and final particles, whose ratio is precisely $\mathcal K_{ij}$.

\subsection{Relativistic Coulomb potential}
\label{sec:rcp}
We consider the relative motion of two particles within the framework of the Klein–Gordon equation  and introduce the electrostatic Coulomb potential through the time component of the covariant derivative via minimal coupling, yielding~\cite{Koshelkin:2025xag}
\begin{eqnarray}
    \left[\left(E_s - \frac{Z_1Z_2\alpha}{r}\right)^2+\vec\nabla^2-\mu_s^2\right]\psi(k, \vec r\,)=0
\end{eqnarray}
where $\mu_s=m_1m_2/\sqrt{s}$ and $E_s=(s-m_1^2-m_2^2)/(2\sqrt{s})=\sqrt{k^2+\mu_s^2}$. Note that, in the non-relativistic limit, $E_s$ reduces to the reduced mass,
$m_1m_2/(m_1+m_2)$, recovering the standard non-relativistic Coulomb
potential. Neglecting terms of order $\mathcal O(\alpha^2)$, 
\begin{equation}
\left[
\vec\nabla^2+k^2 - \frac{2Z_1Z_2E_s\alpha}{r}
\right]
\psi(k,\vec r)=0,
\end{equation}
and introducing the relativistic Sommerfeld parameter $\eta=Z_1Z_2E_s\alpha/k$, the reduced radial $S$-wave function then satisfies
\begin{equation} \label{eq:S19}
u''(k,r)
=
\frac{2\eta k}{r}u(k,r)-k^2u(k,r),
\end{equation}
which is formally identical to its Schrödinger counterpart, with the only
difference being the relativistic definitions of $k$ and $\eta$.
Consequently, the two equations admit the same Coulomb wave-function
solutions when expressed in terms of the appropriate relativistic
kinematics~\cite{Hostler:1963zz}. However, Eq.~\eqref{eq:S19} is not the
differential equation corresponding to the three-dimensional Bethe--Salpeter
equation employed in this work,
\begin{equation}
T^{\rm BS}=V+VG^{\rm BS}T^{\rm BS},
\end{equation}
whose propagator does not coincide with the Klein--Gordon Green's function. Therefore, inserting the momentum-space Coulomb potential directly into the BSE does not reproduce the exact Coulomb solutions.

In fact, Eq.~\eqref{eq:S19} corresponds to the relativistic Lippmann--Schwinger equation
\begin{equation} \label{eq:S21}
T(p',p)
\!=\!
V^{\rm KG}(p',p)
\!+\!\!
\int \!\! \frac{d^3q}{(2\pi)^3}
\frac{
V^{\rm KG}(p',q)\,
T(q,p)
}
{k^2-q^2+i\epsilon},
\end{equation}
with $k^2 = \lambda(s,m_1^2,m_2^2)/4s$, provided that the interaction kernel is given by
\begin{equation}
V_{KG}(p',p)
=
2E_s\,V_C^{\ell=0}(p',p).
\end{equation}
where, in practice, $V_C^{\ell=0}(p,p')$ is the regularized Coulomb potential described in the main text and in Ref.~\cite{Encarnacion:2026iur}.

Comparing Eq.~\eqref{eq:S21} with the BSE, one finds that both become equivalent if the Coulomb kernel is multiplied by the kinematic factor
\begin{equation}
\xi(q;s)
=
2E_s
\frac{2\omega_1(q)\omega_2(q)}
{\omega_1(q) \!+\! \omega_2(q)}
\frac{s \!-\! [\omega_1(q) \!+\! \omega_2(q)]^2}
{k^2-q^2},
\label{eq:coul_fact}
\end{equation}
whose on-shell limit is $\xi(k;s) = 4E_s\sqrt{s}$.
Accordingly, we implement the relativistic  Coulomb interaction through the symmetric kernel
\begin{equation}
V_C^{\ell=0;  {\rm BS}}(p',p)
=
\frac{\xi(p';s)\xi(p;s)}
{\xi(k;s)}
V_C^{\ell=0}(p',p),
\label{eq:coulpotnorm}
\end{equation}
which preserves the symmetry $V_C^{\ell=0;  {\rm BS}}(p',p)=V_C^{\ell=0;  {\rm BS}}(p,p')$. A similar strategy was adopted in Refs.~\cite{Holzenkamp:1989tq,Torres-Rincon:2023qll}, although the kinematic factors employed there do not reproduce exactly the Klein--Gordon kernel of Eq.~\eqref{eq:S19}. The present construction achieves an exact mapping between the two equations while preserving the symmetry of the interaction kernel.

\end{document}